\documentclass[runningheads]{llncs}
\usepackage[T1]{fontenc}
\usepackage{graphicx}
\usepackage{float}
\usepackage{csquotes}
\usepackage{microtype}
\usepackage{tikz}
\usepackage{hyperref}
\usetikzlibrary{svg.path}
\usepackage[framemethod=TikZ]{mdframed}
\usepackage{academicons}
\usepackage{makecell}

\definecolor{orcidlogocol}{HTML}{A6CE39}
\tikzset{
  orcidlogo/.pic={
    \fill[orcidlogocol] svg{M256,128c0,70.7-57.3,128-128,128C57.3,256,0,198.7,0,128C0,57.3,57.3,0,128,0C198.7,0,256,57.3,256,128z};
    \fill[white] svg{M86.3,186.2H70.9V79.1h15.4v48.4V186.2z}
                 svg{M108.9,79.1h41.6c39.6,0,57,28.3,57,53.6c0,27.5-21.5,53.6-56.8,53.6h-41.8V79.1z M124.3,172.4h24.5c34.9,0,42.9-26.5,42.9-39.7c0-21.5-13.7-39.7-43.7-39.7h-23.7V172.4z}
                 svg{M88.7,56.8c0,5.5-4.5,10.1-10.1,10.1c-5.6,0-10.1-4.6-10.1-10.1c0-5.6,4.5-10.1,10.1-10.1C84.2,46.7,88.7,51.3,88.7,56.8z};
  }
}

\newcommand{\OrigHeightRecip}{0.006102}
\newlength{\curXheight}
\DeclareRobustCommand\orcidlink[1]{%
\texorpdfstring{%
\setlength{\curXheight}{\fontcharht\font`X}%
\href{https://orcid.org/#1}{\mbox{%
\begin{tikzpicture}[yscale=-\OrigHeightRecip*\curXheight,
xscale=\OrigHeightRecip*\curXheight,transform shape]
\pic{orcidlogo};
\end{tikzpicture}%
}}}{}}

\newcommand{\tabcaption}[1]{
    \refstepcounter{table}
    \parbox{\linewidth}{\textbf{Table \thetable.} #1}
    \par\vspace{\baselineskip}
}

\newcommand\copyrighttext{%
      \footnotesize \textcopyright~2027 Springer. This document is a preprint. Personal use of this material is permitted.
      Permission from Springer must be obtained for all other uses, in any current or future
      media, including reprinting/republishing this material for advertising or promotional
      purposes, creating new collective works, for resale or redistribution to servers or
      lists, or reuse of any copyrighted component of this work in other works. DOI: not yet available
      }
    \newcommand\copyrightnotice{%
    \begin{tikzpicture}[remember picture,overlay]
    \node[anchor=north,yshift=-12pt] at (current page.north) {\fbox{\parbox{\dimexpr\textwidth-\fboxsep-\fboxrule\relax}{\copyrighttext}}};
    \end{tikzpicture}%
}

\begin{document}
\title{How Use Case Tables Enable\\the Elicitation of Explanation Needs}
\titlerunning{Elicit with Use Case Tables}

%
\author{Lucas Burgath\orcidlink{0009-0003-8025-4480} \and
Jovita Vahlhaus\orcidlink{0009-0003-5013-5501} \and
Jakob Droste\orcidlink{0000-0001-8746-6329} \and
Hannah Deters\orcidlink{0000-0001-9077-7486} \and
Kurt Schneider\orcidlink{0000-0002-7456-8323}}
\authorrunning{Burgath et al.}
%
\institute{Software Engineering Group, Leibniz University Hannover, Germany\\
\email{lucas.burgath@inf.uni-hannover.de}}
\maketitle              

\copyrightnotice
\vspace{-2ex}

\begin{abstract}
\textbf{[Context and motivation]} While software systems get more integrated into people's daily lives – like driver-assistance systems or smart-home applications – they are also becoming increasingly complex. When complex software is used, needs for explanations may arise. At present, explanation needs are usually elicited through the testing of prototypes or existing software. However, detailed documentation of software interactions, such as use case tables, might be a suitable alternative. Since use cases provide early insights into the key interactions within the system, they are appealing to employ as resources for explanation needs.
\textbf{[Question/problem]} As of now, it remains unclear whether use cases allow the elicitation of explanation needs. Or, from a different perspective, whether their utilization for this purpose is desirable.
\textbf{[Principal ideas/results]} We conducted a study in which users perused the description sections of use cases extracted from existing software. While reading through these use cases, they reported any explanation needs they encountered. Our findings indicate that use case steps are sufficient to induce explanation needs in the participants as they read the steps.
\textbf{[Contribution]} This paper explores the novel idea of utilizing use cases for the elicitation of explanation needs and reports on a study that provides insights into the applicability of this concept with stakeholders.

\keywords{Requirements Engineering \and Explainability \and Use Cases}
\end{abstract}

\section{Introduction}



The integration of non-deterministic, opaque system components -- such as artificial intelligence (AI) features -- has put a highlight on explainability as a non-functional requirement (NFR)~\cite{adadi2018peeking,chazette2020explainability}. In essence, explainability describes the ability of a software to provide explanations about itself~\cite{chazette2020explainability,droste2024explanations}. Contemporary research has identified explanations as a driving factor to increase the trustworthiness of complex systems~\cite{adadi2018peeking}. By providing rationale for internal processes and system outputs, explanations can contribute to the understandability of the software, and to the overall user experience (UX)~\cite{deters2024x,droste2024explanations}. However, research in human-computer interaction (HCI) has used explanations to guide interactions and provide learnability, long before explainable AI became an explicit research area~\cite{kieras1984role}. Furthermore, explanations have been researched in software engineering domains such as privacy and security, to contribute to a transparent, trustworthy system design~\cite{brunotte2023privacy,vigano2020explainable}. As such, explanation requirements may concern a variety of system aspects, and remain a relevant topic in both research and practice.

The careful implementation of explanations is a non-trivial task. The need for explanations is highly subjective, and whether an explanation can be considered \enquote{good} depends on the individual that receives it~\cite{sokol2020one}. 
Complex interactions with software often consist of multiple steps, each of which may require explanation. When asked, stakeholders might not be able to recall the exact point at which the need for an explanation occurred. To avoid hypothetical bias and over-reliance on tacit knowledge, the elicitation of explanation needs could be supported with documentation that details the software interactions in question. However, concurrent research into explanation needs is heavily focused on the testing of prototypes or existing systems, and does not yet consider the use of more traditional documentation, such as use cases~\cite{chazette2022can,deters2024pulse}. If the documentation-based elicitation of explanation needs is feasible, it might also enable the consideration of explainability early into the development cycle, when prototypes are not yet available.

To address this research gap, we conducted a study with 13 participants. We wanted to examine whether the elicitation of explanation needs for a software system is feasible using only use case documentation. We prepared six use cases for two different software applications in advance. One software was a specialized application for designing electrical grids (\textit{Envilio}), and the other for general support of organizational work (\textit{Microsoft Teams}). We asked participants to read all twelve use cases and note explanation needs in the description steps. The use cases were displayed and annotated with the help of a specific software tool (\textit{Use Case Tool}), which we provide as supplementary material. Two researchers examined the annotations produced by the participants, identifying 153 explanation needs and achieving substantial agreements. We made two major observations between the applications. Not only did the systems trigger a substantially different number of needs, they also differed in the types of needs that we identified. This suggests that our approach of employing use cases to elicit explanation needs can uncover diverse needs across different software systems and scenarios.

This paper provides two main contributions to requirements engineering for explainable systems:

\begin{itemize}
    \item A preliminary analysis of utilizing use cases for the elicitation of explanation needs -- available on \textit{Zenodo} (for double-blind review at \textit{Figshare}~\cite{SuppMat}).
    \item The \textit{Use Case Tool} used throughout the study to elicit explanation needs with use cases -- available on \textit{GitLab} (for double-blind review at \textit{Figshare}~\cite{SuppMat}).
\end{itemize}

The rest of this paper is structured as follows: First, we present background information and distinguish our idea from existing research in Section~\ref{sec:background-related-work}. In Section~\ref{sec:research-design} we describe our research goals and the procedure we followed. Afterward, we report our results in Section~\ref{sec:results} and discuss them in Section~\ref{sec:discussion}. In Section~\ref{sec:conclusion-future-work} we summarize our findings of this paper and provide an outlook on future work.

\section{Background and Related Work}
\label{sec:background-related-work}
\subsection{Explainability and Explanation Needs}
\label{subsec:bg-explainability}

As an NFR, explainability describes the ability of a software to explain any aspect of itself~\cite{chazette2021exploring,droste2024explanations}. In recent years, research into software explainability has had a strong focus on explainable AI (XAI)~\cite{adadi2018peeking}. Usually, these are use cases in which explanations are used to justify outputs and enable transparent and interpretable system behavior. Outside AI-focused applications, explanations in everyday software use are often needed to guide interactions and provide learnability~\cite{droste2024explanations}. These needs are more in line with traditional human-computer interaction (HCI) research, that used explanations to improve mental model accuracy and user efficiency~\cite{kieras1984role}. Understanding these needs and formalizing them into actionable system requirements is the key objective of requirements engineering for explainable systems.

Whenever a stakeholder wants an explanation for something about a software system, an \textit{explanation need} arises~\cite{droste2024explanations,groen2026stakeholderbedarfe}. Explanation needs are highly subjective, and are influenced by various confounding factors, such as the end user's individual experience with the system, the context of use~\cite{chazette2021exploring,deters2024x,sokol2020one}. Furthermore, design decisions such as the presentation form and timing of an explanation may influence how the overall UX is affected~\cite{brunotte2023privacy,deters2024x}. Taking these factors into consideration, requirements engineers can translate the explanation needs of the stakeholders into formal \textit{explanation requirements}~\cite{groen2026remagazine,groen2026stakeholderbedarfe}.

\subsection{Use Cases}
\label{subsec:bg-use-case}

In requirements engineering, use cases are an established technique for specifying functional requirements and describing system behavior from an external perspective \cite{frattini2025,sinnig2013}. They describe interactions between actors and a system that is required to achieve the actors' goals \cite{pohl2025,sinnig2013}. This serves not only for documentation purposes but also supports communication with stakeholders \cite{dalpiaz2020}.

Use case tables present the detailed flow of a single use case and are an integral component of requirements specifications, such as those following the Volere template \cite{robertson2000}. For the remainder of this paper, we will refer to the structured \textit{use case tables} as use cases. They differ from high-level UML \textit{use case diagrams}, which depict the boundary between the user and the system to clarify which use cases fall inside or outside the system's scope \cite{robertson2000}.

In this paper, we follow Cockburn’s template-based format for use cases, which structures information into defined fields and is available in both textual and tabular representations~\cite{Cockburn1998,frattini2025,pohl2025}. The template starts with the use case name, which briefly states the goal, followed by preconditions that set the expected status when the use case begins, and end conditions that define the system state after a successful outcome. Its core is the main success scenario, outlining all process steps from the trigger to goal completion. Extensions can be attached to specific steps in the main scenario with a condition to describe variations, alternative paths, or exceptional non-success scenarios \cite{Cockburn1998}. We refer to the steps in the main scenario or in the extension sections collectively as use case steps.

\subsection{Related Work}
\label{subsec:related-work}

\paragraph{Explanation Needs:}
Whereas explainability requirements describe the provision of explanations from the system's side, explanation needs address a wish for clarification from the user's perspective~\cite{groen2026remagazine,groen2026stakeholderbedarfe}. As such, requirements engineers must first understand their users' explanation needs, before they can formulate appropriate explainability requirements.
Droste et al.~\cite{droste2024explanations} conducted a survey study with users of everyday software systems, to develop a taxonomy of software explanation needs. The taxonomy divides explanation needs into five main categories with several subcategories. \textbf{Interaction} explanations guide the operation and navigation of software and constituted the highest number of needs. The second most prevalent category of needs concerned \textbf{System Behavior} explanations. This includes explanations of algorithms and unexpected events, similar to XAI explanations. Explanations concerning \textbf{Domain Knowledge} were stated whenever terminology or system-specific elements were unclear. Lastly, some users reported needs to explain \textbf{Privacy \& Security} or the \textbf{User Interface}.


\paragraph{Use Cases:} Concurrent research into explanation needs does not yet consider the use of requirements artifacts, such as use cases or user stories. However, in the closely related field of usability, there is research on the use of these artifacts. Two notable approaches explore how they can be used to identify usability requirements in the early stages of software development.

Hornbæk et al. \cite{hornbaek2007} introduce \textit{Use Case Evaluation (UCE)} in their work, which systematically inspects use cases using guidelines primarily derived from heuristic evaluation to predict potential usability problems. The inspection was conducted by usability experts with no connection to the software.
They compared the use case approach with a conventional think-aloud test and found that 24\% of problem types were identified by both methods. Meanwhile 34\% were identified only through the think-aloud test, whereas 42\% could only be identified through \textit{UCE}.
Qualitative data show that evaluators feel it requires much imagination to visualize a system by its use case.

Another related approach is presented by Oliveira Júnior et al. \cite{Oliveira2020}. They introduce \textit{USARP}, an approach that enriches user stories with personas and usability guidelines to identify and specify usability requirements from the perspective of different user groups. The usability guidelines describe usability mechanisms that influence software design and architecture and can be transformed into functional requirements. These mechanisms include aspects such as system status, feedback, warnings, cancellation, and multilevel help.
An empirical evaluation demonstrates that \textit{USARP} leads to more comprehensive and precise usability requirements and makes the elicitation easier.

Both approaches demonstrate the potential of requirements artifacts to support the identification of usability-related issues or requirements before implementation. However, neither approach directly addresses the systematic identification of explanation needs as the primary objective. Our work builds on this line of research by investigating use case tables as a source for eliciting such needs, with a particular focus on system behavior and interactions.

\section{Research Design}
\label{sec:research-design}

\subsection{Research Goal and Research Questions}

Our goal is to examine the suitability of employing use case tables to elicit explanation needs. To this end, we conducted a user study where we asked participants to read the steps in use case descriptions and annotate those steps for which they need explanations. We pose the following research questions:

\begin{itemize}
    \setlength\itemsep{0.3em}
    \item[] \textbf{RQ1}: To what extent do use case tables allow users to articulate their needs for explanation?
    \item[] \textbf{RQ2}: How unambiguous are explanation needs stated based on use case tables?
    \item[] \textbf{RQ3}: Which categories of explanation needs can be captured with use case tables? 
\end{itemize}

In this paper, we investigate whether use case tables allow users to articulate their explanation needs (RQ1), specifically whether participants are able to formulate their needs for explanation based on the use cases. In addition, we want to investigate whether the use case table approach results in explanation needs that are comprehensible for requirements engineers (RQ2). In other words, we want to find out whether requirements engineers can clearly understand the needs expressed by users. Furthermore, we are aiming to gain initial insights into how well use cases support capturing different categories of explanation needs (RQ3), which allows us to assess for which needs the use of this technique might be worthwhile.

\subsection{Research Process}
\label{subsec:methodology}

\begin{figure}[h]
    \centering
    \includegraphics[width=1\linewidth]{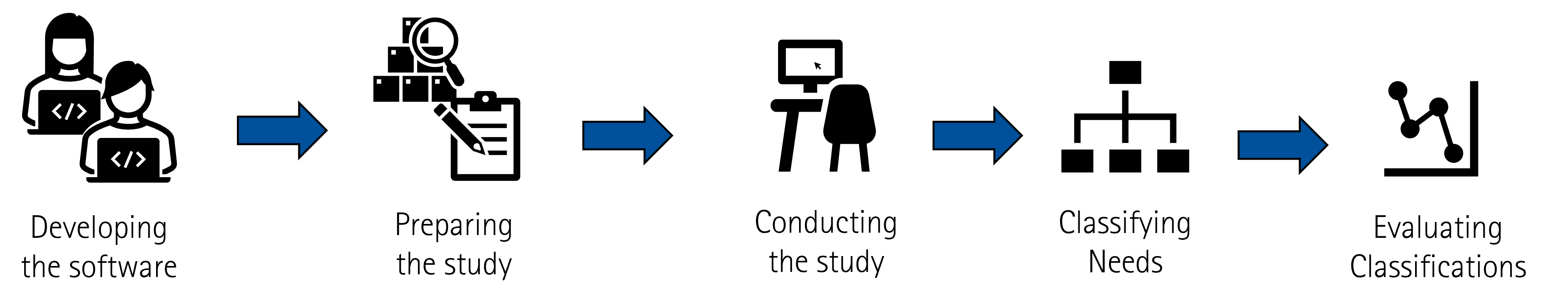}
    \caption{Overview of our methodology.}
    \label{fig:overview-meth}
\end{figure}

The methodology of our research is summarized in Figure~\ref{fig:overview-meth}. At first, we developed a software application -- the \textit{Use Case Tool} -- to help participants report explanation needs during the study. This tool extracts use cases from a software requirements specification and displays them. For each use case step, the tool provides a text field in which users can enter their explanation needs, if applicable. Afterward, we created six use cases for each of the two case applications.  

The participants were asked to peruse the twelve use cases within the tool. They were instructed to focus on the steps in the \textit{Description} and \textit{Extension} sections. While reading through these steps, the participants were tasked with writing down any arising explanation needs. The study was conducted with thirteen participants, each with a technical background. None of the participants needed prior knowledge of explainability, as all received a short introduction to this field.

The stated explanation needs were collected and classified according to the taxonomy of Droste et~al.~\cite{droste2024explanations}. Two coders experienced in the field of explainability independently classified these needs. Any differences in their classifications were resolved with the help of a third coder.

As part of the evaluation, we analyzed the overall occurrence of explanation needs and assessed the unambiguity of the participants' written needs based on the intercoder agreement. Furthermore, we examined the frequency of occurrence of each class across both applications and within each individual application.

\subsection{Participants' Demography}

This study was conducted with 13 employees of the Avacon Netz GmbH from the same site. We decided to recruit participants from a single company. This allowed us to use two case applications for the study that participants are confronted with in their work environment. In this way, we were able to ensure that participants had an intrinsic interest in the software systems. Furthermore, the selected applications are relevant to their daily work, which provides a realistic software use case for the study. The participation in this study was voluntary, and no compensation was provided. All 13 participants completed the entire study, but they were informed they could withdraw at any point. Ten of the participants identified as male (77~\%) and three identified as female (23~\%). The age of the participants ranged between 21 and 59 years. In particular, 46~\% aged between 30 and 39 years old, between 21 and 29 are 31~\% and between the ages of 50 and 59 years are 23~\%. Regarding the highest academic degree achieved by each participant, eight held qualifications as master craftsmen (\enquote{Meister}) or state-certified technicians (\enquote{Techniker}), three held the qualification of a bachelor's degree, one had completed an apprenticeship, and one had a high school diploma (\enquote{Abitur}).

\subsection{Developed Software: Use Case Tool}

\begin{figure}[t]
    \centering
    \includegraphics[width=1\linewidth]{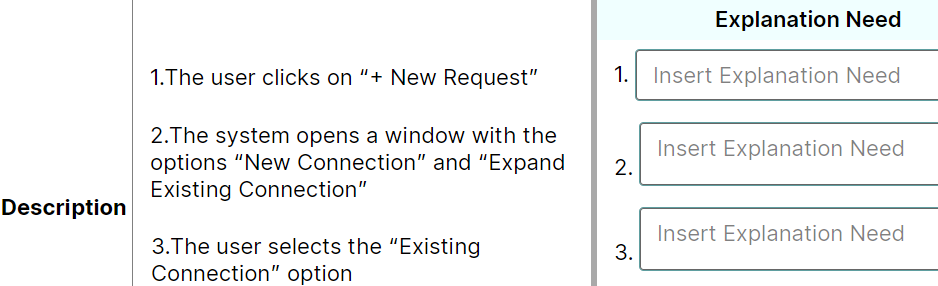}
    \caption{An excerpt of a displayed use case in the tool.}
    \label{fig:use-case-tool}
\end{figure}

The \textit{Use Case Tool} allows users to import a software requirements specification and extract the use cases from it. Additionally, the software can display a use case diagram extracted from the software requirements specification. The tool only extracts use cases that are in tabular format from PDF documents and displays all successfully extracted use cases afterward. Users can select a use case from the list to view the complete use case table. The tool displays all fields included in the original use case table, not just the description and extension steps. Next to each step in the description and the extension sections, the user has an input field to write down explanation needs. This is visually shown in Figure~\ref{fig:use-case-tool}. The annotated use cases can be exported as PDF and CSV files. 

\subsection{Preparing the study}

The study preparation consisted of two activities. Selecting two distinct software applications for the study and creating the needed use cases for these applications. The selected applications are relevant to the participants' work environment which provides a more realistic study scenario. As applications, we picked \textit{Envilio} and \textit{Microsoft Teams} (MSTeams). \textit{Enivilio} supports the design of electrical grids, while \textit{MSTeams} supports organizational tasks. While the latter was well established in the participants' day-to-day work, the former had only recently been introduced at the participating company. 

Following the selection, we created six use cases each for of the two applications. All use cases of \textit{Envilio} contain 67 use case steps in total, while \textit{MSTeams} includes 76 use case steps in total. As orientation, we used Cockburn's use case template~\cite{Cockburn1998}. The drafted use cases ranged from common to rare use cases in both software applications. To match the language of the employees and the developed tool, those use cases were written in German. To ensure the availability of our findings, we provide faithful translations of these use cases and of the stated explanation needs in our supplementary material~\cite{SuppMat}. 

\subsection{Data Collection}
We collected two kinds of data: The annotated explanation needs collected using the Use Case Tool, and the demographic data about the participants through a questionnaire. The study lasted 40 to 50 minutes. 

\subsection{Data Analysis}

The annotated use case tables were exported and manually analyzed. First, two coders experienced in the field of explainability classified the explanation needs according to the taxonomy of Droste et~al.~\cite{droste2024explanations}. The guidelines provided by Droste~et~al.~\cite{Droste2025,Droste2025SupMat} were used to support a more objective classification of explanation needs during coding. The coders could classify any stated explanation need as \enquote{no need}. Examples of unfit needs from our data identified by the coders are remarks like \enquote{very good} or \enquote{This should be possible without a name}. Following the initial coding, a third coder resolved the existing conflicts. Afterward, the intercoder agreement in general, and for each software was calculated. These values allowed the assessment of the clarity and the unambiguity of the collected explanation needs. Lastly, we checked the distribution of needs per category in general and per software. 

\section{Results}
\label{sec:results}

The thirteen participants of our study annotated a total of 165 use case steps in the twelve use cases. A single annotation may include several distinct unclear aspects, which we will refer to as sections. Within the 165 annotations, two coders identified 173 different sections. About 20 (11.56~\%) of these sections did not triggered a need for explanation. In contrast, 153 sections (88.44~\%) triggered a need for explanation. Of the 173 sections we identified, 113 belonged to the use cases of \textit{Envilio}, with 8 sections (7.10~\%) containing no need for explanation. The remaining 60 sections belonged to \textit{MSTeams}, of which 12 (20~\%) contain no need for explanation.

\subsection{Intercoder Agreement}

Both coders classified the sections according to the taxonomy of Droste et~al.~\cite{droste2024explanations}. They achieved an intercoder agreement (ICA) of 76.28~\% and a chance-corrected kappa coefficient of $\kappa = 0.72$, following Brennan and Prediger~\cite{Brennan1981}. This indicates a substantial agreement as defined by Landis and Koch~\cite{LandisKoch1977}. The ICA for both applications together, and per explanation need category, are listed in Table~\ref{tab:ica-per-category}.

\begin{table}
    \centering
    \caption{ICA for sections classified by both coders per explanation need category.}
    \begin{tabular}{|c|c|c|c|c|}
        \hline
        \textbf{Category} & \textbf{\# Sections} & \textbf{Total ICA} & \textbf{ICA MSTeams} & \textbf{ICA Envilio}\\
        \hline
        System Behavior & 130 & 78.46~\% & 80.95~\% & 77.27~\% \\
        \hline
        Interaction & 110 & 75.45~\% & 71.79~\% & 77.46~\%\\
        \hline
        Domain Knowledge & 53 & 71.70~\% & 66.67~\% & 72.73~\%\\
        \hline
        Privacy \& Security & 17 & 88.24~\% & 80.00~\% & 100.00~\%\\
        \hline
        User Interface & 2 & 0~\% & 0~\% & 0~\%\\
        \hline
    \end{tabular}
    \label{tab:ica-per-category}
\end{table}

When considering both applications together, the ICA for \textit{System Behavior}, \textit{Interaction}, and \textit{Domain Knowledge} fall between 70~\% and 80~\%. Across all categories, \textit{Privacy} \& \textit{Security} had the highest ICA at 88~\%, while \textit{User Interface} had the lowest at 0~\%. The values for the first three categories remain relatively similar when examining the applications separately. However, they vary more for \textit{MSTeams} (66~\% - 80~\%) than for \textit{Envilio} (72~\% - 77~\%). The largest difference is for \textit{Privacy} \& \textit{Security}, where the ICA is 80~\% for \textit{MSTeams} and 100~\% for \textit{Envilio}. For \textit{User Interface}, the ICA remains at 0~\% for both applications. The remaining conflicts were resolved by a third coder. If the third coder suggested a category different from those initially selected, a discussion followed to reach a mutual agreement.

\subsection{Categories of Explanation Needs}

The final distribution of explanation needs across categories is listed in Table~\ref{tab:amt-needs}. The distribution across categories per software is presented in Table~\ref{tab:ms-teams-needs} and in Table~\ref{tab:envilio-needs}.

\begin{table}
    \centering
    \caption{Distribution of explanation needs per category after resolving the conflicts.}
    \begin{tabular}{|c|c|}
        \hline
        \textbf{Category} & \textbf{\# Needs} \\
        \hline
        System Behavior & 67 \\
        \hline
        Interaction & 52\\
        \hline
        Domain Knowledge & 23\\
        \hline
        Privacy \& Security & 10 \\
        \hline
        User Interface & 1\\
        \hline
    \end{tabular}
    \label{tab:amt-needs}
\end{table}

Of the 153 explanation needs we found across systems, most were concerned with \textit{System Behavior} (67), followed by \textit{Interaction} (52). A total of 23 needs were found concerning \textit{Domain Knowledge} and ten regarding \textit{Privacy} \& \textit{Security}. One need was assigned to the category \textit{User Interface}. A total of 48 explanation needs were identified for \textit{MSTeams}, compared to 105 for \textit{Envilio}. For \textit{System Behavior} and \textit{Interaction}, the distribution between the applications is similar, with more needs assigned to the former than to the latter category. In contrast, the value for \textit{Domain Knowledge} is 10~\% (5), approximately half the value for \textit{Envilio} at 17~\% (18). The value for \textit{Privacy} \& \textit{Security} is 12~\% (6) for \textit{MSTeams} and one-third of that, 4~\% (4), for \textit{Envilio}. One need was classified under \textit{User Interface} for \textit{MSTeams}, while no needs were assigned to this category for \textit{Envilio}.

\begin{figure}[h]
    \centering
    \begin{minipage}[c]{0.48\textwidth}
        \tabcaption{Distribution of explanation needs per category after resolving the conflicts for \textit{MSTeams}.}
        \centering
        \label{tab:ms-teams-needs}
        \begin{tabular}{|c|c|c|}
            \hline
            \textbf{Category} & \textbf{\# Needs} & \textbf{\%}\\
            \hline
            System Behavior & 20 & 41.7 \\
            \hline
            Interaction & 17 & 35.4\\
            \hline
            Domain Knowledge & 5 & 10.4\\
            \hline
            Privacy \& Security & 6 & 12.5\\
            \hline
            User Interface & 0 & 0\\
            \hline
        \end{tabular}
    \end{minipage}
    \hfill
    \begin{minipage}[c]{0.48\textwidth}
        \centering
        \tabcaption{Distribution of explanation needs per category after resolving the conflicts for \textit{Envilio}.}
        \label{tab:envilio-needs}
        \begin{tabular}{|c|c|c|}
            \hline
            \textbf{Category} & \textbf{\# Needs} & \textbf{\%} \\
            \hline
            System Behavior & 47 & 44,8 \\
            \hline
            Interaction & 35 & 33,3\\
            \hline
            Domain Knowledge & 18 & 17,1\\
            \hline
            Privacy \& Security & 4 & 3,8\\
            \hline
            User Interface & 1 & 1,0\\
            \hline
        \end{tabular}
    \end{minipage}
\end{figure}

\section{Discussion}
\label{sec:discussion}

\subsection{Reflection on the Results}

\paragraph{Varying Numbers of Explanation Needs:} Overall, the number of explanation needs differed between the two examined applications. Notably, roughly twice as many sections were found in the use case tables for \textit{Envilio} (113) compared to \textit{MSTeams} (60). This difference is not attributable to the total number of use case steps, as the use cases for both applications had a comparable number of steps. \textit{MSTeams} has nine more steps overall, underlining that the difference is not driven by the number of steps. Furthermore, the number of sections marked as containing \enquote{no need} for \textit{MSTeams} was approximately one and a half times that for \textit{Envilio}. Conversely, the total number of user-expressed explanation needs was approximately twice as high for \textit{Envilio} as for \textit{MSTeams}. One possible reason for this is that \textit{MSTeams} was designed for a broader stakeholder group and aims to be accessible to a wide range of users, which is why \textit{MSTeams} may be more self-explanatory than \textit{Envilio}. In contrast, \textit{Envilio} was designed for a specific professional purpose and intended for an expert user group. Additionally, this difference could be due to a longer period of prior use. \textit{MSTeams} was already well established in the company, while \textit{Envilio} was still in the introduction phase.

\paragraph{Comparison with other elicitation methods:} With our approach of employing use case tables, each participant identified around twelve needs in a survey-like study lasting around 45 minutes. In our study, participants had access to the taxonomy developed by Droste et al.~\cite{droste2024explanations}. For this reason, we compare our results with studies that also used this taxonomy to ensure that any differences did not arise simply because of the taxonomy’s presence. Droste et al.~\cite{Droste2025} conducted focus groups to identify explanation needs. In that process, approximately ten needs per person were identified in 60 minutes. Obaidi et al.~\cite{Obaidi2025CompareElicitation} examined different approaches for elicitation; in focus groups, participants identified an average of about five needs in approximately 54 minutes. In interviews, participants identified about twelve needs in around 34 minutes. The paper by Deters et al.~\cite{Deters2025Indicators} used an online survey to identify explanation needs. On average, each participant reported approximately two needs. Compared to surveys, our approach took longer, but participants identified significantly more needs. Focus groups took more time and identified fewer explanation needs per participant. Interviews identified a similar number of needs while requiring less time per participant. However, the interviews also required an interviewer to be involved in the process, which is not the case with our approach.

\paragraph{Differences in Explanation Need Categories:} The data further show that \textit{System Behavior} and \textit{Interaction} represent the categories with the highest number of identified explanation needs across both software systems. This aligns with the main purpose of use cases: Describing the interactions between a user and the system.
In contrast to the findings of Droste et~al.~\cite{droste2024explanations}, we observed more explanation needs in \textit{System Behavior} than in \textit{Interaction}. One possible reason for this is that use cases explicitly outline system behavior, often using statements in natural language such as \enquote{the system does...}.
Furthermore, the analysis shows that \textit{Domain Knowledge} has a higher proportion of needs in \textit{Envilio} than in \textit{MSTeams}. 
In contrast, \textit{Privacy} \& \textit{Security} shows a relatively higher proportion of needs in \textit{MSTeams} than in Envilio. This difference may be attributed to \textit{Envilio} being an expert application, whereas \textit{MSTeams} targets a broader audience. Alternatively, users might perceive \textit{MSTeams} as collecting more privacy-related data. 
Use cases seem to lead to few explanation needs within the category \textit{User Interface} because they convey no details about user interaction with the user interface.

\paragraph{Unambiguity of the Explanation Needs:} The sections annotated by the users were assessed by two coders. The coders classified each section into an explanation need category or marked it as \enquote{no need}. The ICA indicates how often the classification of both coders aligned. This can serve as an indirect indicator of unambiguity. A high proportion of ambiguous or unclear statements would likely lead to disagreements between the coders. Disagreements could also result from unclear definitions of the coding categories. However, the coding scheme had already been successfully applied in earlier studies~\cite{droste2026probing}. The substantial agreement achieved, as defined by Landis and Koch~\cite{LandisKoch1977}, indicates that the explanation need statements were relatively clear and unambiguous. It should be noted that both coders were experienced researchers of explainability, which could have influenced the ICA. Due to the low total number of needs found for the category \textit{User Interface} (2), the ICA of 0~\% should be interpreted with caution. Only two explanation needs were assigned to this category across all applications by the two coders. However, neither need was assigned to this category by both coders.

\subsection{Answering the Research Questions}

\subsubsection{RQ1 -- To what extent do use case tables allow users to articulate their needs for explanation?}

Participants were able to note down their explanation needs regarding the use case steps in the use cases of both applications. In total, 173 potential explanation needs were noted by the participants, most of which were classified as valid needs by the coders. The high number of needs identified by the participants, compared with those reported in other studies using different elicitation methods and the same taxonomy, suggests that use case tables helped participants to articulate explanation needs. Participants were able to report explanation needs for both the well-established software (\textit{MSTeams}) and the newly introduced software (\textit{Envilio}). These findings suggest that use case tables can serve as a suitable means of eliciting explanation needs.

\subsubsection{RQ2 -- How unambiguous are explanation needs stated based on use case tables?}

The explanation needs that participants noted alongside the steps in the use case tables were generally interpreted consistently by the coders. The coders achieved substantial agreement when classifying the needs according to the categories defined by Droste et al.~\cite{droste2024explanations}. This suggests that these needs were sufficiently clear and unambiguous to be interpreted consistently. The ICA values ranged from approximately 72~\% to 88~\%, while the \textit{User Interface} category was the main outlier with an ICA of 0~\%. Due to the low number of needs, this ICA does not provide sufficient evidence to draw any reliable conclusions about the ambiguity of this category. Overall, the results suggest that the needs collected via use case tables were generally comprehensible.

\subsubsection{RQ3 -- Which categories of explanation needs can be captured with use case tables?}

The largest number of explanation needs was classified as \textit{System Behavior}, followed by \textit{Interaction}. Both categories were associated with more than 50 needs. This suggests that use case tables are suitable for collecting explanation needs in both categories. More than 20 explanation needs from \textit{Domain Knowledge} were collected using use case tables, suggesting that use case tables can help users report these needs as well. Needs from \textit{Privacy} \& \textit{Security} were also found. Given the relatively small number of needs, the results may depend on the examined application. Finally, only one explanation need was identified for \textit{User Interface}. Therefore, no reliable conclusions can be drawn about the suitability of use case tables to capture explanation needs in this category.

\subsection{Threats to Validity}

We discuss the threats to the scientific validity of our research, following the guidelines of Wohlin et~al.~\cite{Wohlin2012}:

\paragraph{Construct Validity:} These threats relate to whether the study conducted appropriately reflects the theory that was developed. If all participants first viewed the use cases of the same application, a \textit{Learning Bias} could influence our findings. Therefore, half of the participants started with \textit{Envilio}, while the other half started with the use cases of \textit{MSTeams}. All use cases were written by a single researcher, so his writing style could influence the study. To mitigate this influence, another researcher reviewed all use cases. Lastly, a shared understanding of explanation needs among the participants was required. For this reason, we presented and explained the taxonomy of Droste et al.~\cite{droste2024explanations} to all participants beforehand.

\paragraph{Internal Validity:} These threats are associated with whether the results are independent of (unknown) third factors. Existing software was used as the basis for the use cases, which may have helped the participants better imagine them and thus influenced the number of explanation needs they noted down. Additionally, familiarity with one of the software applications could lead to fixed assumptions about how tasks are performed. Our approach may be prone to \textit{false memory bias}~\cite{Seghezzi2027FalseMemoryBias} which might result in listed steps being forgotten by users, leading to more explanation needs. Thus, the explanation needs may arise from this deviation and not from an actual need for an explanation of this step. In particular, \textit{MSTeams} was already well established in the company, unlike \textit{Envilio}. Therefore, a difference in the number of explanation needs could stem from this.

\paragraph{Conclusion Validity:} These threats relate to the statistical correlation between the results and the measurements. A coding procedure is always influenced by the coder's own insights into and understanding of the topic and is therefore subjective. However, the intercoder agreement in our study is substantial, suggesting that our classifications are robust and that other researchers would reach comparable results when coding our data, following the same guidelines.

\paragraph{External Validity:} These are threats related to generalizing the findings beyond the scope of the study. Generalizability is limited because only 13 participants from one domain were included, and only two systems were examined. Additionally, only a few use cases from the two systems were part of the study. These three threats arise from the fact that the study was intended to provide preliminary insights. These threats could be addressed in a more comprehensive study with a larger number of participants, additional applications from various domains, and a broader range of use cases.

\section{Conclusion and Future Work}
\label{sec:conclusion-future-work}

The preliminary insights gained from our study suggest that use cases can help users to articulate explanation needs while reading those use cases. Participants in our study were able to express different types of explanation needs clearly by using use case tables. The results can serve as a take-off point for an in-depth analysis of the applicability and the impact of this procedure within more realistic scenarios. Use cases offer an opportunity to elicit explanation needs in the early phases of software development and could thus allow for the early consideration of unclear user needs within the system.

Whereas we employed use cases from existing applications in this paper, we plan to apply our approach in future research within an ongoing software development project. In this way we want to assess the applicability in a real environment and investigate whether the elicited needs actually arise while using the software. Additionally, we want to investigate whether explanation needs elicited with use cases are similar to those elicited with prototypes or other methods. With these studies we aim to examine the effectiveness of using use cases to elicit explanation needs and the potential benefits of this approach in a software development project.

\begin{credits}

\subsubsection{Data Availability Statement} All materials supporting this study, including the faithfully translated use case tables and explanation needs, the participant demographic data, the explanation needs classified by the two coders, and the repository of the Use Case Tool, are openly available on Figshare at https://figshare.com/s/922811e47aa0dba995be under the CC BY 4.0 License. The replication package contains all materials necessary to reproduce our findings.

\subsubsection{\ackname} This work was funded by the Deutsche Forschungsgemeinschaft (DFG, German Research Foundation) under Grant No.: 470146331, project softXplainer (2026-2029). DeepL was used in parts of this paper to improve the language.

\subsubsection{\discintname}
The authors have no competing interests to declare that are
relevant to the content of this article.
\end{credits}
%
%
%
\bibliographystyle{splncs04}
\bibliography{references}

\end{document}